\documentclass[twocolumn]{jpsj2}

\def\runtitle{Superconducting Gap Modulation in Weak Stripe States }

\def\runauthor{
Masanori \textsc{Ichioka}  and Kazushige \textsc{Machida}
}

\title{Superconducting Gap Modulation in Weak Stripe States }

\author{
Masanori \textsc{Ichioka}\thanks{E-mail address: oka@mp.okayama-u.ac.jp}
 and Kazushige \textsc{Machida}
}

\inst{Department of Physics, Okayama University, Okayama 700-8530}

\recdate{March 15, 2002}

\abst{
The superconducting gap modulation is investigated in the 
presence of a weak stripe structure, using the Bogoliubov-de Gennes 
theory on the two-dimensional Hubbard model with nearest-neighbor 
site pairing interaction. 
We calculate the local density of states and discuss the 
recently observed scanning tunneling spectroscopy spectra 
with four lattice periodicity on 
${\rm Bi_2 Sr_2 Ca Cu_2 O_{8+\delta}}$.  
We also consider the spectral weight in the reciprocal space, 
where the Fermi surface and the superconducting gap are modulated 
by the band folding effect of the stripe structure. 
}

\kword{
stripe state, 
superconducting gap, 
Bogoliubov-de Gennes theory, 
local density of states, 
spectral weight 
}

\begin{document}

%

\sloppy
\maketitle




Recently, much attention has been focused on the stripe state of  underdoped
high-$T_{\rm c}$ cuprates.
The stripe state was proposed to explain the static magnetic incommensurate 
structure observed in elastic neutron scattering experiments on
La$_{2-x}$Sr$_x$CuO$_4$ (LSCO)\cite{Matsuda,Wakimoto,Suzuki} and 
La$_{1.6-x}$Nd$_{0.4}$Sr$_x$CuO$_4$.\cite{TranquadaR,TranquadaNd} 
It is considered that doped holes are localized in the stripe region,
which contributes to one-dimensional (1D) metallic conduction,\cite{Noda}
and the outside region of the stripe is an antiferromagnetic (AF) insulator.
In ${\rm Y Ba_2 Cu_3 O_{7-\delta}}$, 
inelastic neutron scattering experiments reported incommensurate 
fluctuations, which is consistent with the stripe 
concept.\cite{Mook1,Mook2,Arai}

Recently, the incommensurate charge modulation was also reported in   
${\rm Bi_2 Sr_2 Ca Cu_2 O_{8+\delta}}$ (BSCCO).  
Scanning tunneling spectroscopy (STS) revealed modulated quasiparticle 
states with four unit-lattice periodicity surrounding a vortex 
core,\cite{Hoffman}
and recent STS experiments reported that the four unit-lattice periodic 
charge modulation survives even at zero field without a 
vortex.\cite{Howald,Lang} 
In the Fourier component of the checker-board pattern, 
the peak intensitiy at either $\pi(\pm \frac{1}{4},0)$ or 
$\pi(0,\pm \frac{1}{4})$ is dominant
(the lattice constant is taken as unity throughout this paper), 
meaning that we can assume a vertical stripe state and that 
the $x$- and the $y$-direction stripe layers are stacked layer by layer.

Our study is based on the self-consistent Hartree-Fock (HF) theory of 
the Hubbard model. 
It is believed that the stripe concept is valid beyond the HF 
approximation.\cite{Zaanen} 
We can consider the metallic stripe state
under the HF theory if we consider the
realistic Fermi surface topology.\cite{MachidaL,Ichioka}
Once the density of states (DOS) remains at the Fermi energy 
(i.e., metallic state), we can produce superconductivity 
by introducing the pairing interaction, at least as a phenomenological 
model.\cite{IchiokaC,IchiokaV,Martin,Himeda} 
In the previous study on the superconducting state and the vortex state 
in the stripe state, we considered the case where the stripe order is 
stronger than the superconductivity.\cite{IchiokaC,IchiokaV} 
However, experimental results on BSCCO suggest that the stripe order 
is not so great. 
Using the same method as ours, the stripe modulation around the vortex 
or the impurity center has been reproduced for the weaker stripe 
case.\cite{Zhu} 
Therefore, it is valuable to study the STS spectrum for the weak stripe state 
based on this theory. 

In this work,  we investigate how the superconducting gap is modulated 
by the stripe order with four (eight) unit-lattice charge (spin) periodicity 
at zero field. 
We use the Fermi surface topology and the electron 
filling $\sim 0.88$ appropriate to optimally doped BSCCO,\cite{Kordyuk} 
and study the weak stripe state in comparison with the  superconductivity. 
We calculate the local density of states (LDOS), and discuss 
the STS spectrum on BSCCO. 
We also analyze the spectral weight,  
which is observed by angular-resolved photoemission spectroscopy (ARPES),  
to discuss how the Fermi surface and the superconducting gap in the 
reciprocal space are modulated by the weak stripe structure. 

To consider the weak stripe modulation, 
we begin with the conventional Hubbard model on a two-dimensional
square lattice, and introduce the mean field
$ n_{i,\sigma}= \langle a^\dagger_{i,\sigma} a_{i,\sigma} \rangle$
at the $i$-site, where $\sigma$ is a spin index and $i=(i_x,i_y)$. 
We assume a singlet pairing interaction $g_s$ between nearest-neighbor 
(NN) sites, because the coherence length is short in high-$T_{\rm c}$ 
cuprates. 
This type of pairing interaction produces $d$-wave 
superconductivity with the pairing function $\cos k_x-\cos k_y$. 
This gap function is confirmed by the ARPES experiment.\cite{Ding} 
The decomposition to the singlet and the triplet 
components of the pairing interaction is explained in ref. \citen{TakigawaP}. 
Thus, the HF Hamiltonian is given by 
\begin{align}
{\cal H}
&=-\sum_{i,j,\sigma} t_{i,j}a^{\dagger}_{i,\sigma} a_{j,\sigma}
+U\sum_{i,\sigma} n_{i,-\sigma}
a^\dagger_{i,\sigma} a_{i,\sigma} 
\nonumber \\ 
&+g_s\sum_{i,j}
(\Delta^\dagger_{j,i} a_{j,\downarrow} a_{i,\uparrow}
+\Delta_{j,i} a^\dagger_{j,\uparrow} a^\dagger_{i,\downarrow}
) , 
\label{eq:Hamiltonian}
\end{align}
with a creation (annihilation) operator 
$a^{\dagger}_{i,\sigma}$ ($a_{i,\sigma}$). 
For the transfer between nearest, second and third neighbor pairs $(i,j)$, 
$t_{i,j}=t$, $t'$ and $t''$, respectively.  
Then, the dispersion is given by 
$\epsilon(\mib{k})=-2t (\cos k_x+\cos k_y) -4t' \cos k_x \cos k_y 
-2t'' (\cos 2 k_x+\cos 2 k_y) $. 
The singlet pairing potential is defined as 
$\Delta_{j,i}=g_s (\langle a_{j,\downarrow} a_{i,\uparrow} \rangle 
- \langle a_{j,\uparrow} a_{i,\downarrow} \rangle)$. 
When $n_{i, \sigma}$ and $\Delta_{\hat{e},j} (\equiv \Delta_{j,j+\hat{e}})$
have the $N$-site periodicity along the $y$-direction, we can write  
\begin{align} 
& 
n_{j,\sigma} =\sum_{0 \le l <N} 
{\rm e}^{{\rm i}l\mib{Q}\cdot\mib{r}_j} 
n_{\sigma,l\mib{Q}}, \\ 
& 
\Delta_{\hat{e},j}=\sum_{0 \le l <N} 
{\rm e}^{{\rm i}l\mib{Q}\cdot\mib{r}_j} \Delta_{\hat{e},l\mib{Q}} 
\quad (\hat{e}=\hat{x},\ \hat{y}) , 
\end{align} 
with the ordering vector $\mib{Q}=2\pi (\frac{1}{2},\frac{1}{2}-\frac{1}{N})$.
$\hat{x}=(1,0)$ and $\hat{y}=(0,1)$, respectively mean the next sites  
in the $x$- and $y$-directions.  
There is a relation $\Delta_{-\hat{e},j}=\Delta_{\hat{e},j-\hat{e}}$ 
in the singlet pairing. 

After the Bogoliubov transformations  
$a_{j,\uparrow}=\sum_\epsilon ( u_{\epsilon, j} \gamma_\epsilon 
            -{v'}^\ast_{\epsilon, j} {\gamma'}^\dagger_\epsilon )$ 
and $a^\dagger_{j,\downarrow}=\sum_\epsilon (
v_{\epsilon, j} \gamma_\epsilon 
+{u'}^\ast_{\epsilon, j} {\gamma'}^\dagger_\epsilon )$, where 
$\epsilon$ is the label of the eigen states, 
we use the Fourier transformations from 
$u_{\epsilon,j}$ ($v_{\epsilon,j}$) to 
$u_{\epsilon,\mib{k}}$ ($v_{\epsilon,\mib{k}})$. 
We write $\mib{k}=\mib{k}_0+m\mib{Q}$ ($m=0,1,\cdots,N-1$), 
where $\mib{k}_0$ is within the reduced Brillouin zone of size 
$(2 \pi)^2/N$. 
Then, the Bogoliubov-de Gennes (BdG) equation in reciprocal space 
is given by 
\begin{align} 
& 
\sum_{m'}
\left( \begin{array}{cc}
K_{\uparrow,m,m'} & D_{m,m'} \\ D^\dagger_{m,m'} & -K^\ast_{\downarrow,m,m'}
\end{array} \right)
\left( \begin{array}{c} u_{\alpha,\mib{k}_0,m'} \\ v_{\alpha,\mib{k}_0,m'} 
\end{array}\right)
\nonumber \\ 
& 
=E_{\alpha,\mib{k}_0} 
\left( \begin{array}{c} u_{\alpha,\mib{k}_0,m} \\ v_{\alpha,\mib{k}_0,m}
\end{array}\right) ,
\label{eq:BdG1}
\end{align}
where
$K_{\sigma,m,m'}=[\epsilon(\mib{k}_0+m\mib{Q}) - \mu ]\delta_{m,m'} 
+U n_{-\sigma,(m-m')\mib{Q}}$ with chemical potential $\mu$. 
$D_{m,m'}=g_s \sum_{\hat{e}=\hat{x},\hat{y}} 
( {\rm e}^{ {\rm i}(\mib{k}_0+m'\mib{Q})\cdot\hat{e}}
+ {\rm e}^{-{\rm i}(\mib{k}_0+m \mib{Q})\cdot\hat{e}}) 
\Delta_{\hat{e},(m-m')\mib{Q} } $
with  $u_{\epsilon,\mib{k}}=u_{\alpha,\mib{k}_0,m}$ and 
$v_{\epsilon,\mib{k}}=v_{\alpha,\mib{k}_0,m}$. 
The eigen-state $\epsilon$ is labeled by $\mib{k}_0$ and the 
eigen state $\alpha$ of eq. (\ref{eq:BdG1}). 
Then, the wave function of the eigen states is given  by 
\begin{align} 
& 
u_{\alpha,\mib{k}_0}(\mib{r})=N_k^{-1/2}\sum_m u_{\alpha,\mib{k}_0,m}
{\rm e}^{{\rm i}(\mib{k}_0 +m\mib{Q})\cdot\mib{r}}, 
\\ 
& 
v_{\alpha,\mib{k}_0}(\mib{r})=N_k^{-1/2}\sum_m v_{\alpha,\mib{k}_0,m}
{\rm e}^{{\rm i}(\mib{k}_0 +m\mib{Q})\cdot\mib{r}} , 
\end{align} 
where $N_k=\sum_i 1 =\sum_{\mib{k}_0,m} 1$. 
We also obtain the same BdG equation as eq. (\ref{eq:BdG1}) for 
$-{v'}^*_{\alpha,\mib{k}_0,m}$ and ${u'}^\ast_{\alpha,\mib{k}_0,m}$ but with 
the eigen-energy $-E'_{\alpha,\mib{k}_0}$. 
Then, $-{v'}^*_{\alpha,\mib{k}_0,m}$, ${u'}^\ast_{\alpha,\mib{k}_0,m}$, 
$-E'_{\alpha,\mib{k}_0}$ and $\gamma'$ can, respectively, 
correspond to $u_{\alpha,\mib{k}_0,m}$, $v_{\alpha,\mib{k}_0,m}$, 
$E_{\alpha,\mib{k}_0}$ and $\gamma$. 
When there appears spin order, 
$u \ne u'$, $v \ne v'$, and $E \ne E'$, generally. 
The selfconsistent conditions are as follows, 
\begin{align} 
& 
\Delta_{\hat{e},l\mib{Q}} 
=\frac{1}{N_k}\sum_{\alpha,\mib{k}_0,m} [ f(E_{\alpha,\mib{k}_0})\
 {\rm e}^{{\rm i}(\mib{k}_0+(m+l)\mib{Q})\cdot\hat{e}} 
\nonumber \\ 
& \hspace{0.5cm} 
-f(-E_{\alpha,\mib{k}_0}) 
{\rm e}^{-{\rm i}(\mib{k}_0+m\mib{Q})\cdot\hat{e}} ]
v^\ast_{\alpha,\mib{k}_0,m} u_{\alpha,\mib{k}_0,m+l} , 
\\ 
&
n_{\uparrow,l\mib{Q}} 
=\frac{1}{N_k}\sum_{\alpha,\mib{k}_0,m}  
u^\ast_{\alpha,\mib{k}_0,m} u_{\alpha,\mib{k}_0,m+l} 
f(E_{\alpha,\mib{k}_0}), 
\\ 
&  
n_{\downarrow,l\mib{Q}} 
=\frac{1}{N_k}\sum_{\alpha,\mib{k}_0,m}  
v^\ast_{\alpha,\mib{k}_0,m} v_{\alpha,\mib{k}_0,m+l} 
f(-E_{\alpha,\mib{k}_0}) , 
\end{align} 
with the Fermi distribution function $f(E)$.  

The thermal Green's functions  
are given by
$g_{11}(\mib{r},\mib{r}',{\rm i}\omega_n)=\sum_{\mib{k}_0,\alpha}
{u_{\alpha,\mib{k}_0}(\mib{r})u^\ast_{\alpha,\mib{k}_0}(\mib{r}')} / 
{({\rm i}\omega_n -E_{\alpha,\mib{k}_0})} $ and 
$g_{22}(\mib{r},\mib{r}',{\rm i}\omega_n)=\sum_{\mib{k}_0,\alpha}
{v_{\alpha,\mib{k}_0}(\mib{r})v^\ast_{\alpha,\mib{k}_0}(\mib{r}')} / 
{({\rm i}\omega_n -E_{\alpha,\mib{k}_0})}$ 
for up- and down-spin electrons, respectively.~\cite{Takigawa}   
Then, the LDOS 
$N(\mib{r},E)=-\pi^{-1}
{\rm Im}g_{11}(\mib{r},\mib{r},{\rm i}\omega_n\rightarrow E+{\rm i}0^+)
+\pi^{-1}{\rm Im}g_{22}(\mib{r},\mib{r},-{\rm i}\omega_n\rightarrow 
E+{\rm i}0^+)$ 
is reduced to 
\begin{align}
N(\mib{r},E)=
\sum_l {\rm e}^{{\rm i}l\mib{Q}\cdot\mib{r}} N(l\mib{Q},E)  , 
\end{align} 
with the Fourier component 
\begin{align} & 
 N(l\mib{Q},E) = \frac{1}{N_k}\sum_{\mib{k}_0,\alpha,m} 
[u^\ast_{\alpha,\mib{k}_0,m} u_{\alpha,\mib{k}_0,m+l} 
\delta(E-E_{\alpha,\mib{k}_0})
\nonumber \\ & \hspace{2cm} 
+v^\ast_{\alpha,\mib{k}_0,m} v_{\alpha,\mib{k}_0,m+l} 
\delta(E+E_{\alpha,\mib{k}_0})]. 
\end{align} 
Similarly, from the Green's function 
$g(\mib{k},{\rm i}\omega_n)={N_k}^{-1}\sum_{\mib{r},\mib{r}'}
{\rm e}^{-{\rm i}\mib{k}\cdot(\mib{r}-\mib{r}')} 
g(\mib{r},\mib{r}',{\rm i}\omega_n)$ 
in the $\mib{k}$-space, we obtain the spectral weight as~\cite{Ichioka} 
\begin{align} &
A(\mib{k},E) =\sum_{\alpha,\mib{k}_0,m} [
|u_{\alpha,\mib{k}_0,m} |^2\delta(E-E_{\alpha,\mib{k}_0}) 
\nonumber \\ &  \hspace{0.5cm} 
+|v_{\alpha,\mib{k}_0,m} |^2\delta(E+E_{\alpha,\mib{k}_0}) ]
\delta(\mib{k}_0 +m\mib{Q}-\mib{k}) . 
\end{align} 

To reproduce the Fermi surface topology of BSCCO, 
we set $t'=-0.34t$, $t''=0.23t$ and $\mu\sim -0.9t$.
The chemical potential $\mu$ is tuned in order to set the 
spatially averaged electron density $\overline{n_i} \sim 0.88$. 
The essential results of this study do not significantly depend on the 
choice of these parameter values. 
We consider the case of a vertical stripe state with 
the spin periodicity $N=8$.  
The stripe states appear for $U>U_c  \sim 4t$ on the BSCCO-type 
Fermi surface when $g_s=0$. 
The critical $U_c$ is larger than that of the LSCO-type Fermi 
surface.\cite{MachidaL,Ichioka} 
We consider the superconducting state with $g_s=0.8t$  and $T\sim 0$. 
In this case, the stripe state appears for $U > 4.1t$. 
We report the result for $U=4.2t$ and $4.3t$. 
Since the stripe modulation is increased with raising $U$, 
the effect of the stripe is eminent for $U=4.3t$.

%
\begin{figure}[t]
\begin{center}
\includegraphics[width=7.0cm]{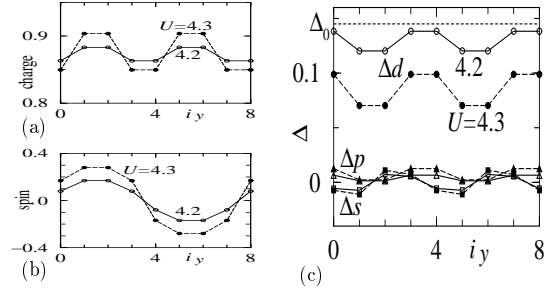}
\end{center}
\caption{
(a) Profile of the charge density,  $n_i$ at the $(0,i_y)$-site.  
$U=4.2t$ (solid line) and $4.3t$ (dashed line). 
(b) Spin density $(-1)^{i_x+i_y} S_i$. 
(c) Superconducting pair potential for 
the $d$-wave $\Delta_{d,i}$ (circle), 
the $s$-wave $\Delta_{d,i}$ (square) and 
the $p_y$-wave $\Delta_{p,i}$ (triangle) components.  
Dotted line shows $\Delta_0$ for the uniform state without stripe ($U=0$). 
}
\label{fig:1}
%
\end{figure}

Figures \ref{fig:1}(a) and \ref{fig:1}(b), respectively, 
show the profiles of the 
charge density $n_i=n_{i \uparrow}+n_{i \downarrow}$ and the spin density 
$S_i=\frac{1}{2}(n_{i \uparrow}-n_{i \downarrow})$. 
The domain wall of the AF is located at the bonds between $i_y$=3 and 4, 
and between  $i_y$=7 and 8. 
Since the bond-centered stripe has lower energy than the site-centered 
stripe in our parameters, we report on the former case. 
Qualitatively, the same results are also obtained for the site-centered 
stripe case. 
At the stripe sites next to the domain wall, 
$1-n_i$ is larger, i.e., doped holes are weakly accumulated. 
Corresponding to this stripe structure, the pair potential of the 
superconductivity is also modulated as shown in Fig. \ref{fig:1}(c), 
where we show the components of $d$-wave 
$\Delta_{d,i}=( \Delta_{ \hat{x},i}+\Delta_{-\hat{x},i}
               -\Delta_{ \hat{y},i}-\Delta_{-\hat{y},i} )/4$,
extended $s$-wave 
$\Delta_{s,i}=( \Delta_{ \hat{x},i} +\Delta_{-\hat{x},i}
               +\Delta_{ \hat{y},i} +\Delta_{-\hat{y},i} )/4$ 
and $p_y$-wave 
$\Delta_{p,i}=( \Delta_{ \hat{y},i} -\Delta_{-\hat{y},i} )/2$. 
The superconductivity is suppressed in total by the stripe formation. 
At the hole-rich stripe site, $\Delta_{d,i}$ has large amplitude. 
When $\Delta_{d,i}$ has a spatial variation, 
small $\Delta_{s,i}$ and $\Delta_{p,i}$ are also induced. 

%
\begin{figure}[t]
\begin{center}
\includegraphics[width=8.0cm]{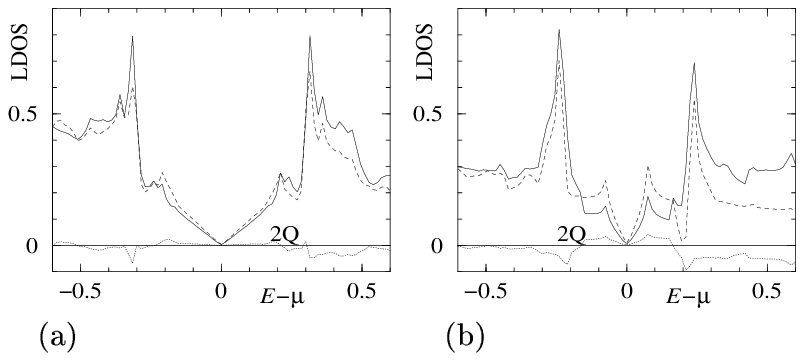}
\end{center}
\caption{
Local density of states $N(\mib{r}_i,E)$ 
at the stripe site $i_y=4$ (solid line) 
and the outside site  $i_y=5$ (dashed line). 
We also show the $2\mib{Q}$-Fourier component 
$N(2\mib{Q},E){\rm e}^{{\rm i}5\pi/4}$ (dotted line).  
$U=4.2t$ (a) and $4.3t$ (b). 
}
\label{fig:2}
%
\end{figure}

Figure \ref{fig:2} shows the LDOS at the stripe site and the outside site. 
In the presence of the stripe structure, there appear two peaks 
inside the gap of the $d$-wave superconductivity. 
These low energy peaks become larger at the outside site with 
weak $\Delta_{d,i}$. 
At the stripe site, the peak at the gap edge becomes sharp.  
We also present the Fourier component 
$N(2\mib{Q},E){\rm e}^{{\rm i}5\pi/4} \propto N(i_y=5,E)-N(i_y=4,E)$, 
showing the difference of the LDOS between the stripe site and the 
outside site. 
The factor ${\rm e}^{{\rm i}5\pi/4}$ is multiplied so that the Fourier 
component becomes real. 
Recently, ref. \citen{Polkovnikov} reported the calculation of 
the $2\mib{Q}$-component assuming carge density wave. 
Since the stripe site gains a large normal state DOS with hole accumulation, 
the DOS above the superconducting gap is larger at the stripe site in 
Fig. \ref{fig:2}. 
When the calculated spectrum is compared with the STS 
spectrum,\cite{Howald,Lang}  
they both show that the $2\mib{Q}$-Fourier component is positive 
(negative) at lower (higher) energies.  
However, the weights of the positive and negative parts of 
the $2\mib{Q}$-component are quantitatively different. 
In the STS spectrum, the weight of the negative $2\mib{Q}$-component at higher 
energies is small, because the peak of the gap edge is broad at the site 
with a large superconducting gap. 
In our calculation, the peak at the gap edge becomes sharp, and the 
weight of the negative $2\mib{Q}$-component is large. 
We give some discussion as follows with respect to this inconsistency. 

(i) 
In order to reproduce the observed STS spectrum, we need to consider 
additional effects of broadening the sharp gap edge at the site with a 
larger superconducting gap, for example, broadening by impurity 
scattering, whose contribution may be larger at the site with a large 
normal state DOS. 

(ii)  
The peak inside the gap (the peak at gap-edge) at the outside (stripe) 
site also appears at the stripe (outside) site, meaning that the 
peak state extends to the adjacent outside (stripe) site. 
However, we see only one peak (higher or lower energy) at each site 
in the STS spectrum. 
This suggests that the coherence of the peak state is restricted to 
within a few sites by the smearing effect of (i), or by other effect 
which confines the peak state to within a few sites. 

(iii) 
We assume that the pairing interaction is constant. 
Then, the site dependence of $\Delta_{d,i}$ comes 
from the site dependence of the DOS. 
We can consider the possibility that the pairing interaction depends on 
the local hole density, having an additional effect on the gap 
inhomogeneity. 
This may be related to the inter-site coherence of the peak state as noted 
in (ii). 

(iv) 
In our calculation, we assume the incommensurate spin order in addition 
to the charge order. 
The STS observes the charge order with the ordering vector $\mib{2Q}$. 
However, the spin order with the ordering vector $\mib{Q}$ has not been 
confirmed experimentally. 
We expect that the ordering with $\mib{Q}$ can be detected by ARPES,  
as discussed later. 

%
\begin{figure}[t]
\begin{center}
\includegraphics[width=8.0cm]{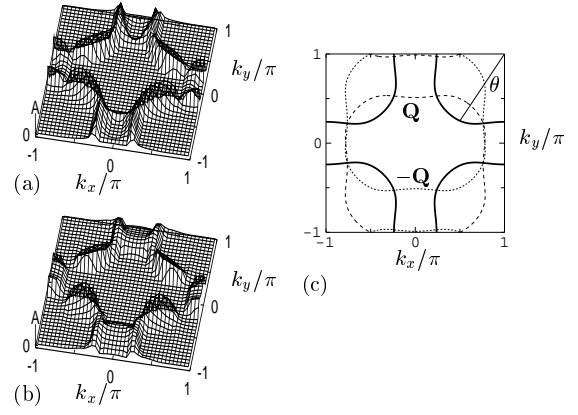}
\end{center}
\caption{
Spectral weight at Fermi energy in the regions  
$|k_x| \le \pi$ and $|k_y| \le \pi$. 
We plot integrated $A(\mib{k},E)$ within the energy range $|E-\mu|<0.5t$. 
$U=4.2t$ (a) and $4.3t$ (b). 
(c) Original Fermi surface (bold line) and $\pm \mib{Q}$-shifted ones 
(dotted and dashed lines).  
}
\label{fig:3}
%
\end{figure}

Next, we discuss how the gap modulation due to the stripe structure 
appears in the reciprocal space.  
The wave number $\mib{k}$-resolved DOS is the spectral weight 
$A(\mib{k},E)$, which is obtained with the same wave function used in the 
LDOS calculation. 
When the system has a periodic structure, 
the band folding effect appears in $A(\mib{k},E)$. 
Figures \ref{fig:3}(a) and \ref{fig:3}(b) show the spectral weight 
at the Fermi energy $\mu$ in the stripe state, integrating $A(\mib{k},E)$ in 
the energy width $\pm0.5t$ around $\mu$. 
In addition to the DOS at the original Fermi surface, 
there appears small DOS near $(\pi,0)$ and $(0,\pi)$ 
in the stripe state. 
As schematically presented in Fig. \ref{fig:3}(c),   
this is due to $\pm\mib{Q}$-shifted Fermi surfaces. 
To induce new DOS by band folding, 
finite DOS is necessary near the Fermi energy in the original dispersion 
$\epsilon(\mib{k})$. 
Around $(\pi,0)$ and $(0,\pi)$, there is large DOS at $E \sim \mu -1.0t$ 
due to the van Hove singularity in our dispersion. 
It is worth noting that the Fermi surface near $(\pi,0)$ and $(0,\pi)$ 
is still the subject of discussion, because additional 
spectral weight appears there.\cite{Chuang,Fretwell,Borisenko}
The spectral weight for the $(\pi,\pi)$-shifted Fermi surface could appear 
due to AF fluctuations as observed in 
the ARPES experiment.\cite{Borisenko} 
If the AF fluctuation is changed to an incommensurate one with 
ordering vector $\mib{Q}$, we expect to observe the spectral weight 
for a $\mib{Q}$-shifted Fermi surface 
instead of a $(\pi,\pi)$-shifted one.  

%
\begin{figure}[t]
\begin{center}
\includegraphics[width=9.0cm]{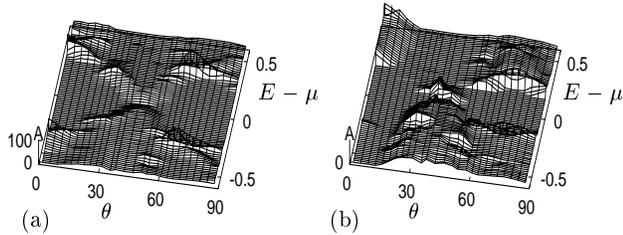}
\end{center}
\caption{
The superconducting gap along the Fermi surface. 
We plot $A(\mib{k},E)$ at $\mib{k}$ along the Fermi surface. 
Wave number $\mib{k}$ is specified by the angle $\theta$ 
from $(\pi,\pi)$, as shown in Fig. \ref{fig:3}(c). 
$U=4.2t$ (a) and $4.3t$ (b). 
}
\label{fig:4}
%
\end{figure}

To discuss the superconducting gap structure in the reciprocal space, 
we show $A(\mib{k},E)$ along the Fermi surface in Fig. \ref{fig:4}. 
The DOS vanishes when $|E-\mu|$ is smaller than the superconducting gap, 
and has a peak at the gap edge. 
In the uniform state without the stripe, the gap is given by 
$|\cos k_x-\cos k_y|$, which has a node at $\theta=45^\circ$.  
In the presence of the stripe, the gap is modulated as shown 
in Fig. \ref{fig:4}. 
Around $\theta=15^\circ$, the Fermi surface DOS vanishes upon the reconnection 
between the original dispersion and  the $\pm\mib{Q}$-shifted dispersions.  
For $\theta>60^\circ$, $\theta$-dependence of the gap becomes flat. 
When the $x$- and the $y$-direction stripe layers are stacked layer by 
layer in the sample, or when there is a domain structure for the $x$- and the 
$y$-direction stripe regions within a layer, 
we observe an overlap of the spectral weights for both direction 
stripe states, i.e., $[A(\mib{k},E)+A(k_x \leftrightarrow k_y,E)]/2$. 

When the stripe modulation becomes strong with increasing $U$ in our system, 
the DOS at the Fermi energy is reconstructed and they form a 
1D Fermi surface at $k_x=\pm \pi/4$, but the DOS vanishes  
near $(\pi/2,\pi/2)$, as suggested in refs. \citen{Ichioka} and \citen{Zhou}. 
In this case, the superconducting gap in the LDOS 
$N(\mib{r}_i,E)$ has a small full gap, because the low-energy state 
near the gap node around $(\pi/2,\pi/2)$  vanishes in reciprocal space. 
This spectrum is seen in Fig. 3 of ref. \citen{IchiokaV}. 
There, the spectrum at the site far from the vortex is almost the same 
as that of the zero-field case without a vortex. 

In summary, we investigate the superconducting gap modulation in the 
presence of a weak stripe structure, using the BdG theory. 
From the same wave functions, the LDOS related to the STS spectrum 
and the spectral weight related to ARPES are calculated. 
In the former, there appear new peaks inside the superconducting gap. 
In the latter, there appears the effect of band folding caused by the 
stripe structure. 
On the basis of our results, we have discussed the STS spectrum with four 
unit-lattice periodicity on BSCCO.\cite{Howald,Lang}

We would like to thank E. Kaneshita, M. Takigawa and N. Nakai 
for valuable discussions. 




\begin{thebibliography}{99}

\bibitem{Matsuda} 
M. Matsuda, M. Fujita,  K. Yamada, R. J. Birgeneau, M. A. Kastner, 
H. Hiraka, Y. Endoh, S. Wakimoto and G. Shirane: 
Phys. Rev. B \textbf{62} (2000) 9148.

\bibitem{Wakimoto}
S. Wakimoto, G. Shirane, Y. Endoh, K. Hirota, S. Ueki, K. Yamada,
R. J. Birgeneau, M. A. Kastner, Y. S. Lee, P. M. Gehring and S. H. Lee: 
Phys. Rev. B \textbf{60} (1999) 769.

\bibitem{Suzuki}
T. Suzuki, T. Goto, K. Chiba, T. Shinoda, T. Fukase,
H. Kimura, K. Yamada, M. Ohashi and Y. Yamaguchi: 
Phys. Rev. B \textbf{57} (1998) R3229. 

\bibitem{TranquadaR}
J. M. Tranquada, B. J. Sternlieb, J. D. Axe, Y. Nakamura and S. Uchida: 
Nature \textbf{375} (1995) 561. 

\bibitem{TranquadaNd}
J. M. Tranquada, J. D. Axe, N. Ichikawa, A. R. Moodenbaugh, Y. Nakamura
and S. Uchida: 
Phys. Rev. Lett. \textbf{78} (1997) 338. 

\bibitem{Noda} 
T. Noda, H. Eisaki and S. Uchida: Science \textbf{286} (1999) 265. 

\bibitem{Mook1}
H. A. Mook, P. Dai, S. M. Hayden, G. Aeppli, T. G. Perring and F. Do\u{g}an: 
Nature \textbf{395} (1998) 580.

\bibitem{Mook2}
H. A. Mook, P. Dai, F. Do\u{g}an and R. D. Hunt: 
Nature \textbf{404} (2000) 729.  

\bibitem{Arai}
M. Arai, T. Nishijima, Y. Endoh, T. Egami, S. Tajima, K. Tomimoto,
Y. Shinohara, M. Takahashi, A. Garrett and S. M. Bennington: 
Phys. Rev. Lett. \textbf{83} (1999) 608.

\bibitem{Hoffman} 
J.E. Hoffman, E.W. Hudson, K.M. Lang, V. Madhavan, 
H. Eisaki, S. Uchida and J.C. Davis: 
Science \textbf{295} (2001) 466. 

\bibitem{Howald} 
C. Howald, H. Eisaki, N. Kaneko and A. Kapitulnik, cond-mat/0201546. 

\bibitem{Lang} 
K.M. Lang, V. Madhavan, J.E. Hoffman, E.W. Hudson, H. Eisaki, S. Uchida 
and J.C. Davis: 
Nature \textbf{415} (2002) 412. 

\bibitem{Zaanen} 
J. Zaanen and A. M. Ole\'{s}: Ann. Phys. \textbf{5} (1996) 224. 

\bibitem{MachidaL}
K. Machida and M. Ichioka: J. Phys. Soc. Jpn. \textbf{68} (1999) 2168.

\bibitem{Ichioka}
M. Ichioka and K. Machida: J. Phys. Soc. Jpn. \textbf{68} (1999) 4020.

\bibitem{IchiokaC}
M. Ichioka and K. Machida: Physica B \textbf{281\&282} (2000) 804. 

\bibitem{IchiokaV}
M. Ichioka, M. Takigawa and K. Machida: 
J. Phys. Soc. Jpn. \textbf{70} (2001) 33. 

\bibitem{Martin} 
I. Martin, G. Ortiz, A. V. Balatsky and A. R. Bishop: 
Europhys. Lett. {\bf 56} (2001) 849. 

\bibitem{Himeda} 
A. Himeda, T. Kato and M. Ogata: 
Phys. Rev. Lett. \textbf{88} (2002) 117001. 

\bibitem{Zhu} 
J-X. Zhu, I. Martin and A.R. Bishop: 
cond-mat/0201519. 

\bibitem{Kordyuk}
A.A. Kordyuk, S.V. Borisenko, M.S. Golden, S. Legner, K.A. Nenkov, 
M. Knupfer, J. Fink, H. Berger, L. Forr\'{o} and R. Follath: 
cond-mat/0201485. 

\bibitem{Ding} 
H. Ding, M.R. Norman, J.C. Campuzano, M. Randeria, A.F. Bellman, T. Yokoya, 
T. Takahashi, T. Mochiku and K. Kadowaki: 
Phys. Rev. B \textbf{54} (1996) 9678. 

\bibitem{TakigawaP} 
M. Takigawa, M. Ichioka and K. Machida: 
Phys. Rev. B \textbf{65} (2002) 014508. 

\bibitem{Takigawa} 
M. Takigawa, M. Ichioka and K. Machida: 
J. Phys. Soc. Jpn. \textbf{69} (2000) 3943. 

\bibitem{Polkovnikov} 
A. Polkovnikov, M. Vojta and S. Sachdev: 
cond-mat/0203176. 

\bibitem{Chuang}
Y.-D. Chuang, A.D. Gromko, D.S. Dessau, Y. Aiura, Y. Yamaguchi, K. Oka, 
A.J. Arko, J. Joyce, H. Eisaki, S. Uchida, K. Nakamura and Y. Ando: 
Phys. Rev. Lett. \textbf{83} (1999) 3717. 

\bibitem{Fretwell} 
H.M. Fretwell, A. Kaminski, J. Mesot, J.C. Campuzano, M.R. Norman, 
M. Randeria, T. Saito, R. Gatt, T. Takahashi and K. Kadowaki: 
Phys. Rev. Lett. \textbf{84} (2000) 4449.  

\bibitem{Borisenko} 
S.V. Borisenko, M.S. Golden, S. Legner, T. Pichler, C. D\"{u}rr, 
M. Knupfer, J. Fink, G. Yang, S. Abell and H. Berger: 
Phys. Rev. Lett. \textbf{84} (2000) 4453.  

\bibitem{Zhou}
X.J. Zhou, P. Bogdanov, S.A. Kellar, T. Noda, H. Eisaki, S. Uchida, 
Z. Hussain and Z.-X. Shen: Science \textbf{286} (1999) 268. 


\end{thebibliography}
\end{document}